\documentclass[journal]{IEEEtran}
\usepackage{ifpdf}
\usepackage{cite}
\ifCLASSINFOpdf
\usepackage[pdftex]{graphicx}
\usepackage{amsmath}
\usepackage{algorithmic}
\usepackage{array}
\usepackage{caption}
\usepackage{subcaption}
\usepackage{cleveref}
\usepackage{mhchem}
\expandafter\def\expandafter\quote\expandafter{\quote\small}

\usepackage{fixltx2e}
\usepackage{stfloats}
\usepackage{url}
\begin{document}
\title{An Open Source Representation for the NYS Electric Grid to Support Power Grid and Market Transition Studies}

\author{~M. Vivienne Liu,
        ~Bo Yuan,~\IEEEmembership{Student Member,~IEEE},
        ~Zongjie Wang,
        ~Jeffrey A. Sward,
        ~K. Max Zhang,
        and~C. Lindsay Anderson,~\IEEEmembership{Senior Member,~IEEE}
\thanks{M. V. Liu is with the Field of Systems Engineering, Cornell University, Ithaca, NY, 14853, USA (e-mail: ml2589@cornell.edu)}
\thanks{C. L. Anderson is with the Field of Systems Engineering and Cornell Energy Systems Institute, Cornell University, Ithaca, NY, 14853, USA (e-mail: cla28@cornell.edu)}
\thanks{B. Yuan, J. A.Sward and K. M. Zhang are with Sibley School of Mechanical and Aerospace Engineering, Cornell University, Ithaca, NY, 14853, USA (e-mail: by276@cornell.edu; jas983@cornell.edu; kz33@cornell.edu).}
\thanks{Z. Wang is with the Department of Electrical and Computer Engineering, University of Connecticut, Storrs, 06268 CT, USA (e-mail: zongjie.wang@uconn.edu).}}


\maketitle

\begin{abstract}
Under the increasing need to decarbonize energy systems, there is coupled acceleration in connection of distributed and intermittent renewable resources in power grids. To support this transition, researchers and other stakeholders are embarking on detailed studies and analyses of the evolution of this complex system, which require a validated representation of the essential characteristics of the power grid that is accurate for a specific region of interest. For example, the Climate Leadership and Community Protection Act (CLCPA) in New York State (NYS) sets ambitious targets for the transformation of the energy system, opening many interesting research and analysis questions. To provide a platform for these analyses, this paper presents an overview of the current NYS power grid and develops an open-source\footnote{https://github.com/AndersonEnergyLab-Cornell/NYgrid} baseline model using publicly available data. The proposed model is validated with real data for power flow and Locational Marginal Prices (LMPs), demonstrating the feasibility, functionality, and consistency of the model. 
The model is easily adjustable and customizable for various analyses of future configurations and scenarios that require spatiotemporal information about the NYS power grid with data access to all the available historical data and serves as a practical system for general methods and algorithms testing. 
\end{abstract}


\IEEEpeerreviewmaketitle

\section{Introduction} \label{sec:Introduction}

\IEEEPARstart{P}{ower} grid models are crucial for power system development, planning and operational analysis. Meineche et al. reviewed existing steady-state power grids test cases and datasets that are publicly available and summarized the background information with possible use cases \cite{meinecke2020review}. Most of the grid datasets reviewed are intended to be a generic testbed for steady-state studies such as OPF and reliability analysis \cite{wang2021progressive, wang2019simulation}. However, power grids exhibit distinct features and behaviors in  different regions. The significant investments in wind, solar, storage units and other renewables only increase the regional specificity due to the availability of resources under various geographical and climate features \cite{chen2021shocks}. For example, New York State (NYS) has passed the Climate Leadership and Community Protect Act (CLCPA) \cite{clcpa}, which targets 9 GW of offshore wind, 3 GW of energy storage and 6 GW of solar by 2035, 2030 and 2025, respectively. To evaluate the feasibility of such an aggressive plan and to optimally allocate and operate a combination of highly distributed and variable resources, it is crucial to perform analyses on a grid model that can reflect the principal physical characteristics of the NYS grid. However, specific power grid data is often unavailable due to security and economic concerns \cite{birchfield2016grid,medjroubi2017open}, which is a challenge for region-specific research. Therefore, in this paper, we propose a representative model for the NYS power grid that captures key physical characteristics of the real grid, in its current form, with publicly available data only. The model is intended to represent the basic network topology, load allocation, generation condition for different fuel types, and the connecting proxies with neighboring grids. The intent is for this model to serve as a baseline model for future analyses, supporting  higher renewable penetration reliably and effectively. In addition to the contribution for NYS, the baseline model also provides a generic practical testbed with high network topology and data coherency, closely representing a real-world power grid, for testing of methods and algorithms.

Several studies have focused on building region-specific power grid models around the world. A research tool for the European interconnected system was developed by Zhou and Bialek to study the effects of cross-border trades \cite{zhou2005approximate}. The authors collected generation and demand data and created a transmission line database using GIS information. The model was then validated with Optimal Power Flow (OPF) and Power Transfer Distribution Factor (PTDF) analysis. Nutcheon and Bialek later improved the European model and further validated it with Power Flow (PF) analysis \cite{hutcheon2013updated}. A dataset to describe the largest transmission network of Australia was constructed for open access by Xenophon and Hill~\cite{xenophon2018open} to support studies that aim to address engineering, economic and environmental challenges of the existing electricity system. Thorough validation was carried out to test the completeness, consistency and functionality of the network. Similar work has been conducted for the Southeast Asian region by Ahmed et al.~\cite{ahmed2018modelling} to help future development scenarios of the Association of Southeast Asian Nations (ASEAN) network. In the US, the Federal Energy Regulatory Commission (FERC) \cite{krall2012rto} and the Western Electricity Coordination Council (WECC) \cite{price2011reduced} test system provide large-scale testbeds for PJM area and the Western Electricity grid, respectively. An 8-zone ISO-NE test system was developed to assist fast-execution simulation of the current ISO New England (NE) system in~\cite{krishnamurthy20158}. These studies based on publicly available data are able to represent principal physical characteristics of the real power grids, and most are validated use simulation against historical data. As NYS embarks on a period of aggressive transition, it is critical to provide an accessible NYS-centric grid model to support analyses. 

Several academic networks have been created specifically for NYS in the past two decades. Allen et al \cite{allen2008combined}, developed a 36-bus network reduced from the FERC 715 case for the Northeastern power grid with a focus on NY and New England (NE). The model preserves the external interfaces with NYS and fits generator bidding curves for OPF analysis but  \begin{quote}
it is intended to be a test system for algorithms and software, as opposed to a study system for the Northeastern region. \emph{(Allen et al., 2008)}
\end{quote} 
Howard et al. proposed a transshipment model \cite{transship} to study the GHG emission of NYS, where the generators of different types and demands were carefully modeled for the 11 load zones (A-K) in NYS. However, with the focus on GHG emission factors, the model is not capable of OPF analysis and thus has limited capability for long-term planning, reliability assessment or electric market-related studies \cite{howard2017current}. A NYS academic model was presented by Burchett et al. to study the effects of steam unit ramp rates under different renewable penetration scenarios \cite{burchett2018investigation}. In this study, a 68-bus network for the Northeast Power Coordinating Council (NPCC) area was reorganized to represent four load zones in NYS and one external load zone in New England. Coded in MATPOWER \cite{matpower}, the model can easily perform OPF analysis, but the transmission lines and the topology of the network were designed to describe the connectivity for the whole NPCC area. Without updating the transmission line parameters, the model is not able to accurately represent the physical characteristics of the NYS network. Additionally, the connections with other surrounding power grids are ignored in this representation. In addition to the papers mentioned above, representative networks can be found from the 2019 Congestion Assessment and Resource Integration Studies (CARIS) report \cite{Independent2018} and the 2020 Reliability Needs Assessment (RNA) report \cite{Assessment2020} of the New York Independent System Operator (NYISO), where the transmission topology and connectivity for load zones and external areas are provided. However, the transmission line parameters are not given. While these NYISO examples~(\cite{Independent2018, Assessment2020}) inform the the construction of the NYS baseline network, they are not directly usable for analysis.

In this paper, we propose a baseline model for NYS, developed with publicly available data to represent the principal physical characteristics of the NYS power grid. The proposed network is reduced and modified from the NPCC 140-bus model \cite{chow1991user} to take advantage of the geospatial information, meeting the desire for future distributed renewable energy integration research with a focus on New York State. Generation profiles (for various fuel types) and load data are collected and compiled for a full year to perform DC-PF-related analysis. The cost curves for different types of generators are fitted with dynamic fuel price input to support DC-OPF studies. We focus on DC-OPF due to the lack of reactive power data, but it is worth noting that AC-OPF could potentially be performed on the model if reactive power data are carefully estimated. We validate our PF simulation results against historical data on major interface power flows collected from NYISO \cite{nyisodata} so that when the baseline model is used for a diverse set of applications in the future, the result reliably represents the NYS system. The PF simulation from our model shows promising results for most of the hours in the year 2019. To further demonstrate that the proposed baseline model represents the real NYS network well, we conduct DC-OPF analysis for winter and summer seasons and compare the simulated zonal LMPs with historical data. To summarize, the proposed model has the following improvements and features relative to existing models: 

\begin{enumerate}
    \item \textit{The network represents the NYS power grid with geographical information, intending to be a research tool for renewable integration impact in the region.} 
    \item \textit{The model can backtrack through the publicly available historical data according to users' preferences and needs\footnote{for more detail see repo at https://github.com/AndersonEnergyLab-Cornell/NYgrid}. }
    \item \textit{The model is validated with real data for PF and OPF analysis to test its fidelity using year 2019 as an example.}
    \item \textit{The model and a sample dataset are open-source and available at: https://github.com/AndersonEnergyLab-Cornell/NYgrid}
\end{enumerate}

The rest of the paper is organized as follows: Section~\ref{sec:NYGrid} provides an overview of the NYS power grid. The data sets collected and processed for the baseline model are described in Section~\ref{sec:dataset}. Section~\ref{sec:Method} presents the equivalent reduction of the original NPCC 140-bus model and integrates the processed data with the reduced network. We validate our model with PF comparison to historical data in Section\ref{sec:validation} and show demonstrative OPF cases for summer and winter 2019 in Section~\ref{sec:testcase}. Section~\ref{sec:conclusion} concludes the paper with discussion of the benefits and drawbacks of the model and future working directions.

\section{Overview of The NYS Power Grid}~\label{sec:NYGrid}
\begin{figure}
     \centering
     \begin{subfigure}[b]{0.35\columnwidth}
         \centering
         \includegraphics[width=0.99\columnwidth,trim = {0cm 2cm 0cm 0cm},clip]{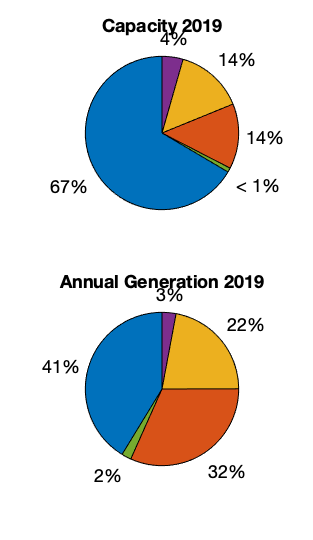}
         \caption{Capacity and Annual Generation in 2019}
         \label{fig: CapGen}
     \end{subfigure}
     \hfill
     \begin{subfigure}[b]{0.63\columnwidth}
         \centering
         \includegraphics[width=0.99\columnwidth,trim = {1.5cm 0cm 2cm 0cm},clip]{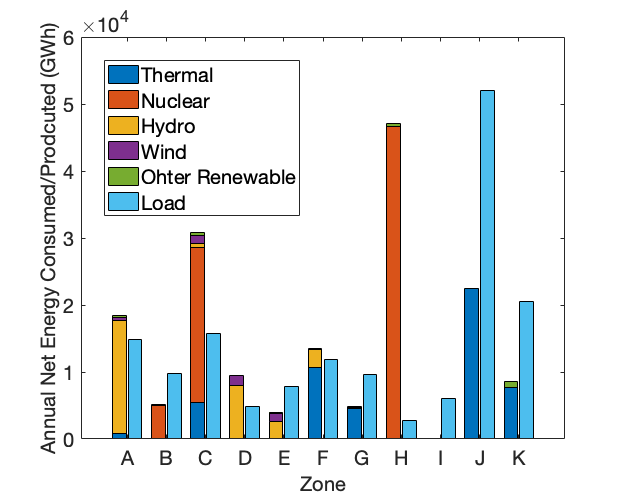}
         \caption{Composition of annual generation and demand for each load zone in 2019}
         \label{fig: fuelcompositon}
     \end{subfigure}
     
        \caption{Overview for the NYS power grid in 2019}
        \label{fig: overview}
\end{figure}

The NYS power grid has eleven load zones labeled A-K, connecting to four neighboring power grids. The composition of the existing generating sources, annual generation and demand for each load zone can be found in \cite{NewYorkIndependentSystemOperator2019,goldbook2020} and are summarized in Figure \ref{fig: overview}. As shown in Figure \ref{fig: CapGen}, nuclear and hydro sources contributed over 55\% of 2019 annual generation, residing in zones A, D and zones B, C, H, respectively. Wind and other renewable sources contributed less than $6\%$ of the total generation, where the wind output ($3\%$) is exclusively in upstate zones (A-E) and other renewables ($<3\%$) are distributed approximately evenly in all zones. Thermal generators mostly reside in downstate zones (F, G, J and K), which leads to complicated bidding situations as we describe in detail in Section \ref{sec:testcase}. Another characteristic of the NYS grid is the lack of balance in generation and demand spatially. The upstate zones (A-E) have generation capability that exceeds loads, while the downstate zones (F-K) have more load than generation capacity. As a result, power generally flows from upstate to downstate. Due to congestion and lack of generation capacity, zonal marginal prices are usually higher for downstate zones (F-K) than upstate zones(A-E). 

The NYS grid is connected to four neighbouring areas with ten proxies: Hydro Quebec (two proxies), Ontario (IESO, one proxy), ISO-New England (three proxies) and PJM (four proxies). Four of the proxies are controllable HVDC lines: 1385 line (ISO-NE - NY), Cross Sound Cable (ISO-NE - NY), Neptune (PJM - NY) and HTP (PJM - NY). One of the proxies, Linden VFT (PJM - NY), is a controllable AC line, and the rest are uncontrollable AC lines. 

NYISO directs the operation of the NYS power system, supplying power to serve load and maintaining safety and reliability of the grid \cite{NYISO_UG}. NYISO provides open access to the NYS transmission system for the market participants and facilitates a two-settlement market clearing process: a Day-Ahead Market (DAM) and a Real-Time Market (RTM). In the DAM, market participants sell or buy electricity one day before the operating day to avoid price volatility. A set of generators are scheduled to be available for dispatch in each hour of the next day and a set of Load Serving Entities (LSEs) are scheduled to buy a specific quantity of power at the day-ahead price. In the RTM, market participants trade electricity during the course of the operating day to balance the differences between day-ahead commitments and the realized real-time demand and production\cite{NYISO_UG}. In both market clearing processes, Locational Based Marginal Prices (LBMPs) are calculated through unit commitment and economic dispatch using market participants' bidding prices. Electricity suppliers and LSEs may trade energy directly in the market at LBMP, or be party to bilateral contracts.

\section{Dataset Processing} \label{sec:dataset}
In this section, we introduce the data collected and processed to inform the grid model. We first present the NPCC 140-bus network that represents the Northeast Power Coordinating Council area. Then an example is provided to illustrate the use of 2019 demand and generation data to update the operating conditions for the proposed baseline model. Lastly, we describe the interface power flow and LBMP data that is used to validate the model. 

\subsection{Network dataset} \label{subsec:Networkdata}
\begin{figure}
  \begin{center}
  \includegraphics[width=2.8in]{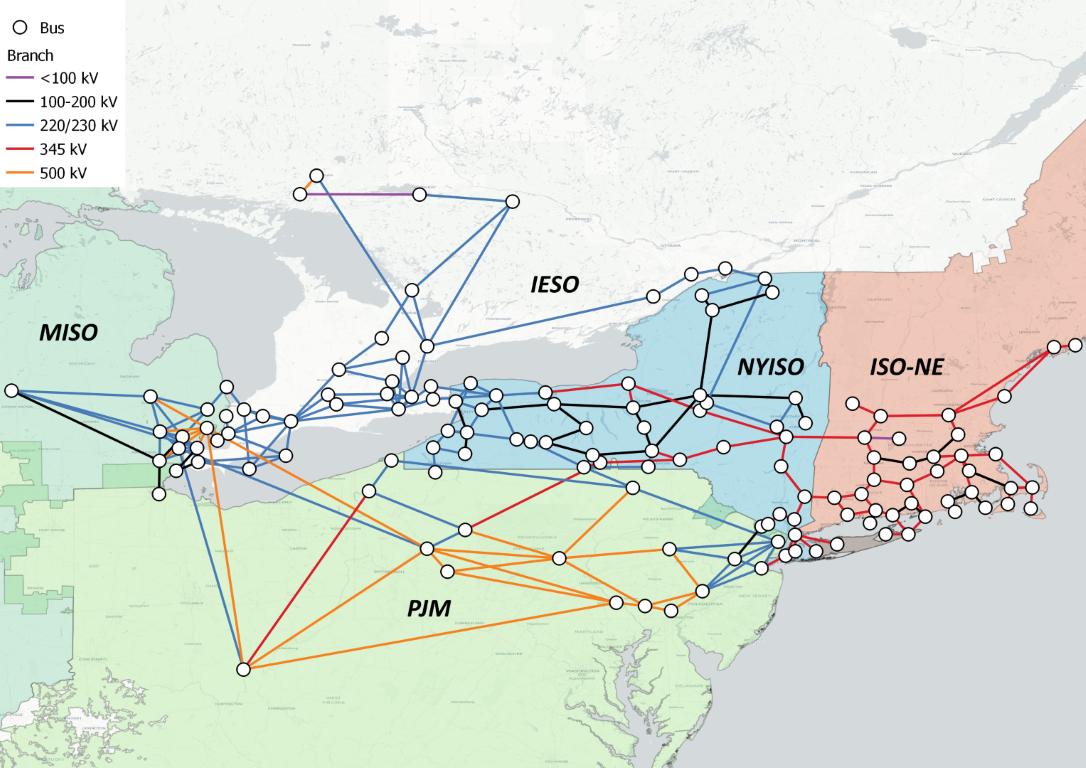}
  \caption{The NPCC 140-bus system represents the backbone transmission system of northeast region of the Eastern Interconnection: NYISO (Blue), NE-ISO (Red), PJM (Light-green), MISO (Darker-green), IESO (White).}\label{fig:npcc140}
    \vspace{-20pt}
  \end{center}
\end{figure}

The NPCC 140-bus system is a power system test case from the Power System Toolbox \cite{chow1991user}.  As shown in Figure \ref{fig:npcc140}, it represents the backbone transmission lines of the Nnortheast region of the Eastern Interconnection \cite{ju2015simulation}, which has full or partial representations of five ISOs: NYISO (full), NE ISO (partial), PJM ISO (partial), MISO (partial), IESO (partial) with 140 buses, 48 generators and 233 transmission lines. A baseline load and generation operation condition is given with the network data for power flow analysis. More importantly, the locations of all the buses are available, which provides the ability to incorporate spatially correlated information of conventional and renewable generation resources. Such a network provides a full representation of the NYS area and preserves the major interfaces with neighboring grids, providing the ability to model the interactions between NYS and other ISOs/RTOs. 

\subsection{NYS Generation Data}\label{subsec:NYGendata}
NYISO \cite{nyisodata} provides aggregated hourly generation for different fuel types, which we will refer to as fuel mix data within the paper. However, the hourly profiles for individual generators, which are crucial for validating the proposed baseline model and performing PF-based analysis, are not available. Thus, we compile and estimate the generation profiles for all the available generators from NYISO's 2019 Load \& Capacity Data Report  \cite{NewYorkIndependentSystemOperator2019}, where the locations, capacities and aggregated annual generations are provided for all the generators. Cost curves with fluctuating fuel prices\cite{Independent2018} are fitted to support OPF-based research.

\subsubsection{Thermal Generators}
Information regarding thermal generators in NYS is compiled from NYISO, Regional Greenhouse Gas Emission reports, and the Energy Information Administration (EIA) as described below. Specifically, 227 fossil fuel generators are identified in NYISO’s 2019 Load \& Capacity Data Report \cite{NewYorkIndependentSystemOperator2019}\footnote{note that~\cite{NewYorkIndependentSystemOperator2019} excludes some small generators with zero annual generation.}

Hourly generation and emission data of 140 fossil fuel generators are available from the Regional Greenhouse Gas Initiative (RGGI) \cite{rggi}, which is the first mandatory market-based program in the United States to reduce greenhouse gas emissions. 
RGGI requires fossil fuel power plants with capacity greater than 25 MW to obtain an allowance for each ton of \ce{CO2} they emit annually \cite{rggicc}. The total nameplate capacity of the RGGI tracked fossil fuel generators is 26,782 MW, which is 92.4\% of the total nameplate capacity of all fossil fuel generators. The other 87 fossil fuel generators are mostly combustion turbines in New York City (zone J) and Long Island (zone K) that are too small to be tracked by RGGI.
Thus, thermal generators in zone J and K are scaled slightly to compensate for the untracked 7.6\% of thermal generation in NYS. We use the hourly generation data to calculate generator parameters, such as heat rate, maximum power, minimum power, and ramp rate~\cite{howard2017current}. For the fossil fuel generators with data recorded by RGGI, their heat rate curves are derived through a linear regression of the hourly heat input data and hourly power output data. A linear approximation of heat rate relation is chosen because it can describe the data with $R^2$ greater than 0.9 for most generators. Note that poor fit is exhibited for some small gas turbines that lack sufficient data as they only operate a few hours in a year. The maximum power is defined as the maximum hourly power output recorded in the RGGI database for 2019. The minimum power is defined as the 5th percentile of all the recorded hourly power output in the same year. The hourly ramp rate is defined as the absolute value of maximum change in hourly power output over the year ~\cite{howard2017current}. For downward ramp rates, we excluded the data points for which power output in the next hour is zero, to prevent error introduced by generator shutdown.

For small generators with insufficient generation data recorded in 2019, mostly small gas turbines, maximum and minimum power output are set to their nameplate capacity and zero, respectively. For these generators, hourly ramp rate is set equal to maximum power output for each, since small gas turbines have high ramp rate and can reach full output in one hour. Standard heat rate curves are defined using EIA's Electric Power Annual Report according to their unit type and fuel type\cite{eia}. A fossil fuel generator's cost curve is defined as its heat rate curve multiplied by the fuel price ignoring the variable operation and maintenance cost, which is relatively small. Time-varying fuel costs for natural gas, fuel oil, and coal are obtained from EIA \cite{eia}.

\subsubsection{Nuclear Generators}
Individual hourly nuclear generation profiles are not publicly available, so profiles are estimated from the daily capacity factor obtained from U.S Nuclear Regulatory Commission \cite{usnuclear} with the following assumption given the limited data accessibility:
\begin{itemize}
    \item Hourly capacity factor is the same as the daily capacity factor for each individual generator in a day.\footnote{Users can supply their own hourly nuclear capacity factors if desired}
\end{itemize}
Thus, the hourly output for each nuclear power plant is: 
\begin{equation}
    P_{i,t}^{N} = \Bar{P}_{i}^{N} \times cf_{i,t}^{N}
\end{equation}
Where $\Bar{P}_{i}^{N}$ is the capacity of nuclear generator $i$. $P_{i,t}^{N}$ and $cf_{i,t}^{N}$ are the hourly output and capacity factor for nuclear generator $i$ at hour $t$, respectively. 
The ten minute ramp rate and marginal cost for nuclear generators are assumed to be 10\% of the generator capacity and 1-3 \$/MWh given nuclear generators typically serve as base load and operate at 100\% capacity if not in maintenance. 

\subsubsection{Renewable Generators}\label{subsubsec:renewable}
New York State has three categories of renewable resources in the fuel mix data for the real-time dispatch (every five minutes). The installed capacity and annual generation of 2019 are shown in Figure \ref{fig: overview}.  We average the 5-minute interval fuel mix data into hourly and estimate generation profiles for hydro, wind and other renewables (PV included) as follows.

\begin{figure}
  \begin{center}
  \includegraphics[width=0.8\columnwidth]{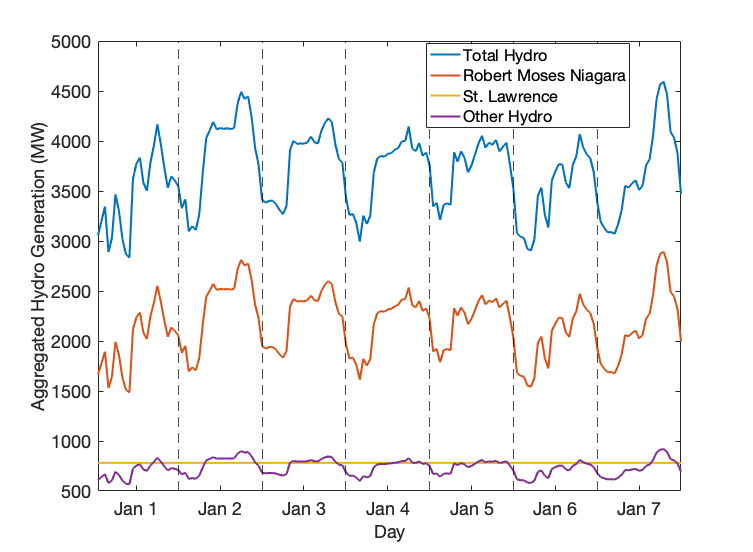}
  \caption{Aggregated hydro generation with strong diurnal pattern for Jan 1- Jan 7, 2019.}\label{fig: hydrogen}
  \vspace{-20pt}
  \end{center}
\end{figure}

\textbf{Hydro Generation}
The total generation of the 347 hydro plants in NYS has a strong diurnal pattern, shown in Figure \ref{fig: hydrogen} by the blue line. The two largest hydro power plants: Robert Moses Niagara (in load zone A) and St. Lawrence (in load zone D), consistently contributed roughly 80\% of hydro generation in 2019\cite{eia923}. The strong diurnal pattern is dominated by the Robert Moses Niagara power plant which bids much lower during night hours (an upper limit of 1800 MW) than during daylight hours (upper limit of ~2600MW)\cite{nyisodata}. Whereas, the St. Lawrence power plant prioritizes regulating water level and thus will not vary much in terms of hourly generation. modeling hydro power takes economic, environmental, agricultural and political factors into consideration \cite{doering2021diagnosing,turner2020data}. As a result, the following simplifications and assumptions are made to estimate hourly individual hydro profiles without over-complicating the model: 
\begin{itemize}
    \item St. Lawrence (STL) hydro plant has a monthly capacity factor, as this is the highest resolution of capacity factor publicly available from the Energy Information Administration (Form EIA-923)\cite{eia923}.
    \item Excluding the two largest hydro generators in zone A and D, the remaining smaller hydro plants generate 20\% of total hourly hydro generation.  
    \item Robert Moses Niagara (RMN) hydro plant generates the remaining portion of hydro power and thus dominates the diurnal pattern.
\end{itemize}

The hourly hydro generation can then be calculated as follows:
\begin{equation}
     P_{i,t}^{H} = 
\begin{cases}
    cf_{STL,m(t)}^{H} \Bar{P}_{i}^{H}  ,
    & \text{if } i =  \text{STL}\\
    0.8P_{total,t}^{H} - P_{STL,t}^{H},              & \text{if } i =  \text{RMN}\\
   0.2P _{total,t}^{H} \dfrac{\Bar{P_i}^{H}}{\sum\limits_{{i\notin\text{STL, RMN}}}\Bar{P_i}^{H}} ,
    & \text{otherwise} 
\end{cases}
\end{equation}

Where, $P_{i,t}^{H}$ is the hourly generation for hydro $i$ at time $t$, $cf_{STL,m(t)}^{H}$ denotes the capacity factor of month $m(t)$ for St. Lawrence hydro and $\Bar{P_i^H}$ is the capacity for hydro $i$. The decomposed generations of RMN, STL and all other hydro plants are shown in red, yellow and purple lines in Fiugre \ref{fig: hydrogen}, respectively. The 10 minute ramping rates for hydro are usually quite large and assumed to be 90\% of its capacity. Because hydro generators do not have fuel costs, their marginal cost is the variable operations and maintenance cost, which is assumed to be a uniform random variable within range 1-10\$/MWh to prioritize the dispatch of hydro generation. 

\textbf{Wind Generation} There are 24 wind farms scattered across zones A-E. From the Form EIA-923 \cite{eia923}, the percentage of wind generation contributed by each zone changes very little by month, which means on the aggregate level, the wind farms are highly correlated. Therefore, we assume perfect correlation of all the wind farms and calculate the individual output as follows:
\begin{equation}
     P_{i,t}^{W} = \dfrac{\Bar{P_i}^{W}}{\sum_{i}\Bar{P_i}^{W}}P_{total,t}^{W} 
\end{equation}
Where $P_{i,t}^{W}$ is the  generation output of wind farm $i$ at hour $t$, $\Bar{P_i}^{W}$is the installed capacity for wind farm $i$ and $P_{total,t}^{W}$ is the total wind generation of hour $t$ from the fuel mix data.
Wind generation is assumed to be non-dispatchable and is modeled as negative load in our simulations, with a marginal cost assumed to be zero. It worth noting that wind resources could be easily converted to dispatchable resources in future settings. 

\textbf{Other Renewables} We identify 46 generators powered by other renewable resources, including solar, methane, refuse, and wood. Most of these generators have capacity under 25MW and monthly aggregated output is not available. Because the total capacity of other renewables is quite small and they are scattered in nine different load zones, their impact to the  grid overall is negligible. Output of these small generators are therefore allocated as a capacity weighted fraction of the output from ``other renewables", as follows:
\begin{equation}
     P_{i,t}^{OR} = \dfrac{\Bar{P_i}^{OR}}{\sum_{i}\Bar{P_i}^{OR}}P_{total,t}^{OR} 
\end{equation}
Where $P_{i,t}^{OR}$ is the generation of unit $i$ at hour $t$, $\Bar{P_i}^{OR}$ is the capacity for other renewable generator $i$ and $P_{total,t}^{OR}$ is the total generation of other renewables at hour $t$ from the fuel mix data. Similar to wind, we assume these renewables are non-dispatchable and model them as negative load in the system with zero marginal cost. 

\subsection{NYS Load Data} \label{subsec:NYloaddata}
Real time and day-ahead load data for each load zone can be accessed from NYISO's database \cite{nyisodata}. Both 5-minute interval and hourly load data are available for the real time market. Hourly integrated load data is collected to complement the real-time hourly generation profile as described in Section \ref{subsec:NYGendata}. 

\subsection{Interface Power Flow Data}\label{subsec:Interfacedata}
To validate the proposed model, major interface flows observed by NYISO are collected for 2019\cite{nyisodata} . The dataset contains 19 major interface flows recorded roughly every five minutes, where seven measurements are for internal zonal interfaces and the remaining (12) represent external trades. From NYISO’s Reliability Needs Assessment (RNA) report \cite{Assessment2020}, the internal (zonal) interfaces are identified and matched with NYISO labels as summarized in Table~\ref{tab:interface} to support discussion of results in Section~\ref{sec:validation}. It is worth noting that the positive limits are dynamically determined by NYISO and exhibit some variability.

\begin{table}[h]
\centering
\caption {\label{tab:interface} Internal Interfaces and Relevant Load Zones}
\resizebox{\columnwidth}{!}{
\begin{tabular}{llll}
\hline
\hline
 Interface Name & Corresponding Load Zones  & Positive Limit 
 Range (MWh) & Negative Limit(MWh)\\   \hline
 Dysinger East & $A \rightarrow B$ & 2100 - 3150 & -9999\\
 West Central & $B \rightarrow C$ & 9999 - 9999 & -9999\\
 Moses South & $D \rightarrow E$ & 500 - 3150 & -1600\\
 Total East & $E \rightarrow F$,$E \rightarrow G$ & 6800 - 6800 & -9999\\
 Central East & $E \rightarrow F$ & 1270 - 2985 & -9999  \\
 UPNY ConEd & $G \rightarrow H$ & 4800 - 5700 & -9999\\
 SPR/DUN-South & $I \rightarrow J$ & 1500 - 4600 & -9999\\
\hline
\hline
\end{tabular}}
\end{table}
\vspace{-10pt}
\subsection{Local Marginal Price Data}\label{subsec:lmpdata}
The real time zonal locational based marginal price (LBMP) data is collected for 2019, which is the weighted average price of energy across all the nodes in each zone. The NYISO dataset contains 15 regional prices for every 5-minute interval. Four of the LBMPs represent external nodal interfaces with PJM, ISO-NE, Ontario (IESO) and Hydro Quebec (HQ). The remaining 11 zonal prices are for the load zones within NYS. We compare the LBMPs with the marginal generation cost of each zone derived by DC-OPF in Section \ref{sec:testcase}. For simplicity, the LBMP is referred to as LMP throughout the paper. The LMP recorded by NYISO is used to represent marginal costs for generators in the external areas for the simulation study.

\section{Comprehensive Network Modeling} \label{sec:Method}
Equivalent network reduction is frequently used to effectively represent a larger system while preserving internal system characteristics and significantly reducing the number of buses in the external system. WARD \cite{WARD} and REI \cite{REI} or their variants are the commonly used methods for equivalent network reduction. Comparison between methods show that the WARD-type equivalents have higher accuracy than REI for real power flow, effectively representing the eliminated PQ buses, and can be easily incorporated into power flow problems. In addition, this method can track real-time variations in operating points by updating only power injection ranges. Therefore, to reduce computational complexity and simplify the neighboring regions, the modified Ward-type equivalent \cite{shi2012optimal} is applied to the  NPCC 140-bus system to reduce the number of buses in areas outside of NYS. 

Before applying the network equivalent reduction algorithm, which is sensitive to the load and generation condition of the network, we make the following simplifications and updates to NPCC 140-bus model:

\begin{enumerate}
\item \textit{Simplify interactions between external areas:}\\
    Three transmission lines connecting PJM and MISO are removed so that the external areas only connect to NYS, making it easier to control the external interfacial flows while performing PF validation.
    
\item \textit{Update outdated load and generation conditions in NYS:}\\
    To encourage the convergence of the PF solutions, the NYISO-observed zonal load data within NYS is allocated to each bus proportional to the baseline load conditions provided by the NPCC 140-bus system. The generators are allocated to the nearest bus within their load zone. Four buses are converted from PQ buses to PV buses (39, 77, 45, 62) and four from PV buses to PQ buses (53, 54, 68, 72) accordingly. The slack bus is moved from bus 78 to bus 74 due to the absence of generation in Zone I.
    
\item \textit{Update the hourly load conditions in external areas:}\\
    The load of each external area $x$ is scaled up by a factor of $\mathbf{\frac{D_x}{D^\prime_x}}$ for each bus, where $\mathbf{D_x}$ and $\mathbf{D^\prime_x}$ are the updated and original load of area $x$. The reasoning is two-fold: first, the loads within NYS are updated proportionally to the original load condition, so maintaining proportional load condition in external areas will maintain consistency with the baseline case for the whole system. Second, the external areas will be reduced to a few highly aggregated virtual buses with virtual load and generators by the network reduction, so spatially refined updates to external zones will not be represented in the final reduction.
    
\item \textit{Maintain interface flows between external areas and NYS:}\\
    Given the interface flow data described in Section \ref{subsec:Interfacedata}, and the removal of connections between external areas, each area is now connected only to NYS and the following equations hold for power balance:
\begin{equation}
    \mathbf{P_{PJM}} = \mathbf{D_{PJM}} + \mathbf{l_{PJM \rightarrow NY}}
\end{equation}
\begin{equation}
    \mathbf{P_{NE}} = \mathbf{D_{NE}} + \mathbf{l_{NE \rightarrow NY}}
\end{equation}
\begin{equation}
    \mathbf{P_{IESO}} = \mathbf{D_{IESO}} + \mathbf{l_{IESO \rightarrow NY}}
\end{equation}
where $\mathbf{P_{x}}$ and $\mathbf{D_{x}}$ are the total generation and demand for area $x$, $\mathbf{l_{PJM \rightarrow NY}},\mathbf{l_{NE \rightarrow NY}},\mathbf{l_{IESO \rightarrow NY}}$ are the interface flows from each area to NY. 

\item \textit{Update hourly generation conditions in external areas:}\\
The generation scaling factor for each area can be calculated as:
\begin{equation}
\alpha_{
x} = \mathbf{P_{x}}/\mathbf{P_{x}^\prime}
\end{equation}
where $\mathbf{P_{x}^\prime}$ is the baseline external area generation provided by the NPCC140-bus network. Then the generation condition for generator $i$ in the external area $x$ can be updated by:
\begin{equation}
    \mathbf{P_{x,i}} = \alpha_{x}\mathbf{P_{x,i}^\prime}
\end{equation}
where $\mathbf{P_{x,i}^\prime}$ is the baseline generation condition for generator $i$ provided by the NPCC 140-bus network.

\item \textit{Include Hydro Quebec in the NPCC 140-bus model:}\\
The original NPCC 140-bus system did not include an interface with HQ. Given the importance of this external zone, a hydro generator is added in zone D, with generation output equal to the interface flow $\mathbf{l_{HQ \rightarrow NY}}$.
\end{enumerate}

After the modifications described above, the original NPCC 140-bus system is updated with load and generation profile and suitable for the PF-based equivalent reduction. 

To perform the network reduction, we select 57 buses to be preserved including 46 buses representing the NYS area and nine boundary buses that are proxies for external areas. In addition, two additional buses are preserved: bus 132 is held to minimize the PF error of equivalent network, and bus 21 is preserved to allow preservation of HVDC lines to the reduced network later. The validation of the reduced model against the original NPCC 140-bus model can be found in section \ref{subsec:val_reducednet}. 
The reduced network has very similar zonal connections to the shipment model in the NYISO RNA report \cite{Assessment2020} with the exception of the missing connection between E and G. A $345 \text{ kV}$ AC transmission line is added to represent the Marcy South line and the reactance parameter (p.u) is estimate based on the following equation. 
\begin{equation}
    X = \dfrac{1.5L_{Marcy}X_L}{Z_{base}}
\end{equation}
Where $L_{Marcy}$ is the distance between the two substations connected by the Marcy South transmission line, $Z_{base} = \frac{V_{base}^2}{S_{base}}$ and $X_L = 0.4~ohms/mile$ \cite{tranparams}. We adopt a factor of 1.5 to approximate the real length of the transmission line \cite{xenophon2018open}, which would be longer than the direct distance between substations. The HVDC lines are added to the reduced network, reflecting the controllable interfaces by connecting buses (in original index): 21 – 80 (CSC+NPX1385), 124-79 (Neptune) 125-81 (VFT), and 125-81 (HTP).

\begin{figure*} [h!]
  \begin{center}
  \includegraphics[width=0.8\textwidth]{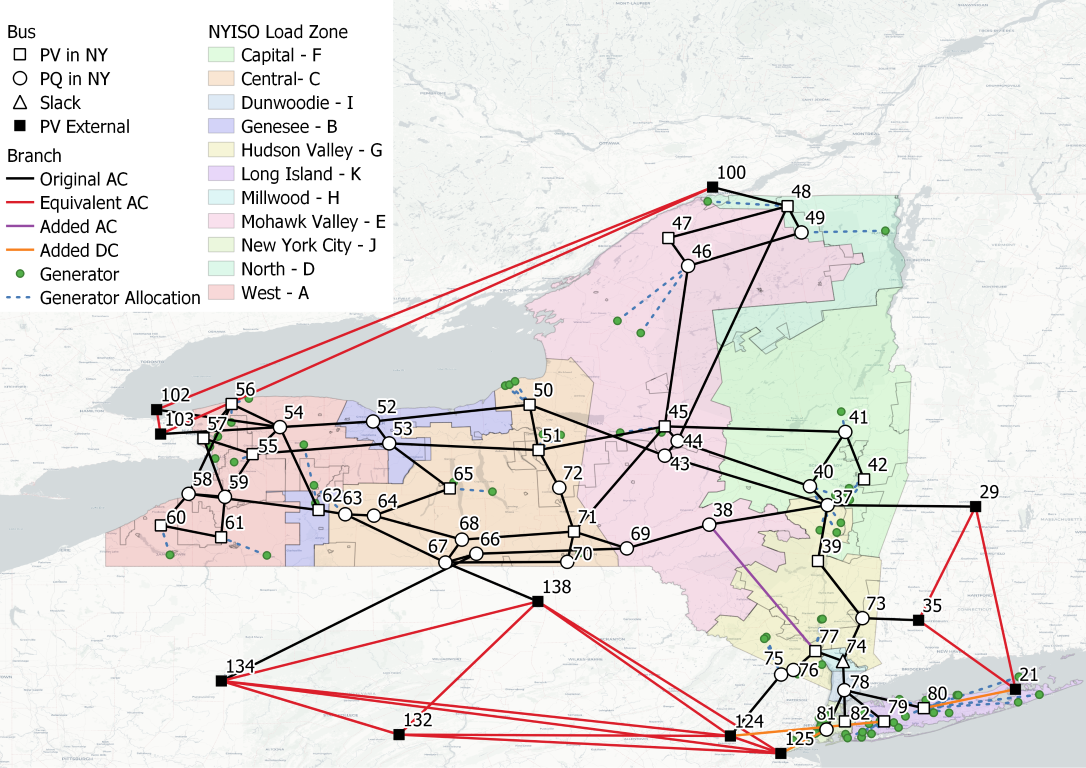}
  \caption{Reduced and Modified Network for NYS}\label{fig: Reducednet}
  \vspace{-10pt}
  \end{center}
\end{figure*}

The final form of the reduced network is shown in Figure \ref{fig: Reducednet} where PV, PQ and slack buses in NYS are shown as empty squares, circles and triangles, respectively. The external proxies are shown as solid squares. The solid black lines in the NYS area maintain their original parameters, whereas the solid red lines represent the equivalent virtual lines in the external areas, to support the same PF solution. The orange lines represent the added HVDC lines and the purple line indicates the added AC line between zone E and G. The thermal generator (green dots) allocations are indicated by the dashed blue lines. 

\section{Model Validation}\label{sec:validation}
In this section, the power flow (PF) consistency of our reduced model is first tested by comparison to the original network. Feasibility and consistency are then tested with the real-world data processed for 2019, followed by discussion of the impact of hydro on model performance. 

\subsection{Validation of Consistency with Original Model} \label{subsec:val_reducednet}
The original NPCC 140-bus model represents the back bone of the northeast power grids. Thus, after updating the generator and load profiles for the network, it is critical to verify that the PF solutions of the reduced model are sufficiently similar to the original network on the retained branches. We compare the DC-PF solutions on each branch in the NYS area. All the absolute errors are in the negligible ($10^{-9}$) range which supports the claim that the reduced model has equivalent DC-PF solution to the original model, after simplification of the external areas.

\subsection{Feasibility and Consistency with Real Data}\label{subsec:val_feasibility}
\begin{figure} [h!]
  \begin{center}
  \includegraphics[width=0.85\columnwidth, trim={0cm 0cm 1cm 1cm},clip]{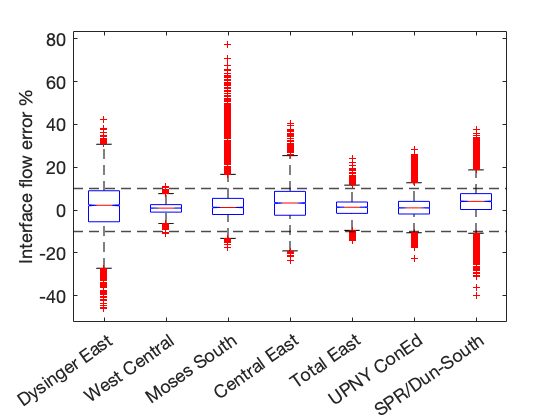}

  \caption{Box plot for simulated power flow percentage error of seven major interfaces in 2019}\label{fig: nonviolin}
  \vspace{-10pt}
  \end{center}
\end{figure}

A primary objective of our study is to provide a baseline model for NYS, which captures the characteristics of the real power grid without revealing or requiring confidential information. PF is one of the basic and critical foundations for many applications. Therefore, PF analysis is conducted for all hours in 2019 and results are compared with historical records on the major interface flows.
Figure \ref{fig: nonviolin} shows the box plots of the percentage power flow error, which is calculated as:\\
\begin{equation}
    \epsilon = \dfrac{pf_i^{r} - pf_i^{s}}{L_{i,MVA}}
\end{equation}
where $pf_i^r$ is the real PF recorded and $pf_i^s$ is the simulated PF for interface $i$, $L_{i,MVA}$ is MVA rating of interface $i$.\\
 All the simulation cases have feasible solutions for PF, which indicates high feasibility of our proposed model given real-world data. The interquartile ranges are within $\pm10\%$ and have high densities.
 The 95\% interval of each interface are mostly within $\pm15\%$ except for Dysinger East (A-B) and Moses South (D-E), which tend to deviate due to hydro impacts. It is worth noting that the outliers of Dysinger East and Moses South exhibit some symmetry in that the outliers that are negative for Dysinger East are positive for Moses South. This is potentially caused by the assumption that the St. Lawrence hydro has constant monthly output, which ignores the temporal variability of hydro outputs. The positive outlier of Moses South indicates that extra hydro generation has been allocated to St. Lawrence, and Robert Moses Niagara generation is lower than it should be.  In addition to hydro impacts, other possible contributors of inaccuracies include: (a) estimation error of wind and other renewable generation, (b) averaging of real-time data: as these data are not recorded at precise 5-minutes intervals, the mean of each hour is not exactly the hourly output/power flow. c) a small portion of thermal generation profiles ($ \approx 7.6\%$ of the total thermal capacity) are not known: the mismatched thermal generators are allocated to zones J and K as previously mentioned, which potentially caused  bias in the UPNY-conEd (G-H) and SPR/SUN-South (I-J) interfaces. 
Considering the factors above, the limitations resulting from data accessibility, and the relatively small errors, this model adequately characterizes the underlying spatiotemporal structure for the real NYS power grid.  

\section{Test Case}\label{sec:testcase}
In this section, the model is further tested, with the cost curves fitted to generators, to illustrate the capability of this model for simulating accurate  zonal LMPs. As stated in Subsection \ref{subsec:lmpdata},  real-time market LMP data is collected as a reference. However, LMPs depend on several factors that cannot be easily modeled in a universal baseline model. Such factors include, but are not limited to, the bidding strategy of different types of generators, long-term contracts with external areas, or generators that are used for multiple purposes (e.g. hydro or steam turbines), unplanned outages,  maintenance and special planning, or controlled regulation due to extreme weather conditions. While this precludes to the comparison of LMPs on a one-to-one basis, Pearson correlation coefficients can be used to evaluate the similarity between simulated and real LMPs arising from common historical conditions. To test the fidelity of our model, we run DC-OPF for winter (Dec-Jan) and summer (Jun-Aug) seasons. The correlation coefficients are first analyzed between derived zonal LMPs with historical LMP records, followed by a more comprehensive analysis for particular ``outliers”.
\begin{figure}[h!]
  \begin{center}
  \includegraphics[width=0.99\columnwidth,trim = {0cm 0 0cm 0},clip]{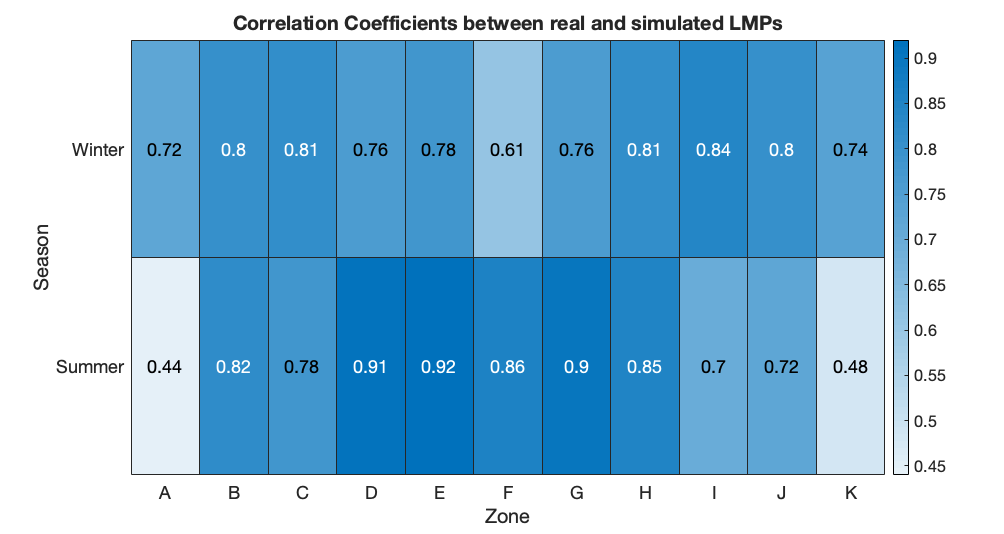}
  \vspace{-10pt}
  \caption{Correlation Coefficients of Historical and Simulated LMPs by season and zone}\label{fig: heatmap}
  \vspace{-10pt}
  \end{center}
\end{figure}
\subsection{Correlation Analysis}\label{subsec:testwinter}
The correlation coefficients between the simulated and historical LMPs for each zone are shown in Figure \ref{fig: heatmap}. One extremely large point which has $-1300 \$/MWh $ in zone D is removed for winter and eight large prices were removed zone A, I and K where the LMPs are larger than $400\$/MWh$ for summer. This ensures that the correlation coefficients are not underestimated by a few significant outliers.
During the winter season, correlations are strong (around 0.8) in most zones, except zones A, D, F and K. Zone A and D are significantly impacted by large hydro plants, as discussed in Section \ref{sec:validation}, where a refined model for dispatch and bidding strategies is out of the scope of this paper. Zone F contains 10 thermal generators, which are also potentially impacted by the the bidding strategies in some cases. For example, The Bethlehem Energy Center facility ran for 8438 out of 8760 hours in 2019. However, if we assume it bids at the cost curve fitted (with high $R^2 = 0.98$) in the DC-OPF model, it won't be dispatched for most of the times due to its high cost compared to other generators. One hypothesis is that the Bethlehem Energy Center was bidding at a lower price because it can make profit by the steam it generates for other purposes or it has a long-term contract with LSEs. Zone K has a similar but even more complicated situation with 72 thermal generators. This means even just a small portion of generators are not bidding at their energy cost curve or have other byproducts such as steam, the OPF model cannot dispatch them correctly with cost curves only. Reserve resources can be another reason to cause the diverge in zones F and K.

The correlation coefficients are higher for summer months, with the exception of zones A, I, J and K. Referring to Figure \ref{fig: overview}, these are zones with significant hydro or thermal generation. Summer peak demand is much higher at 31 GW, relative to 23 GW for winter. Through analysis of bidding curves, it is worth noting that the bidding curve which seems to represent Robert Moses Niagara has much larger variation in summer than in winter. The impact of this  is an increased error in the simple OPF model, leading to lower correlation coefficients in summer for zone A. Conversely, the bidding curve for St. Lawrence is quite stable in the $0-10\$/MW$ range which is close to assumed parameter, resulting in a better correlation between observed and simulated values in zone D. Zone I has no generating facilities and thus its price is driven by neighboring zones H, J and K. J and K both have very complicated thermal compositions and as the reserve units might bid at extremely high prices with higher demand in summer, the simulated LMPs for J and K diverge more from the real ones, causing a lower correlation coefficient than other zones.

To conclude, the baseline model can capture the statistical trends for the real LMPs for most zones with correlation coefficients at around 0.8 for winter and 0.9 for summer. A more complicated bidding model that takes into consideration the unique features for hydro and thermal units could potentially improve the results for the baseline model. We demonstrate the potential with ``Outlier Analysis" . 

\subsection{Outlier Analysis}
\begin{figure}
     \centering
     \begin{subfigure}[b]{0.49\columnwidth}
         \centering
         \includegraphics[width=\textwidth,trim = {0.5cm 0 1cm 0},clip]{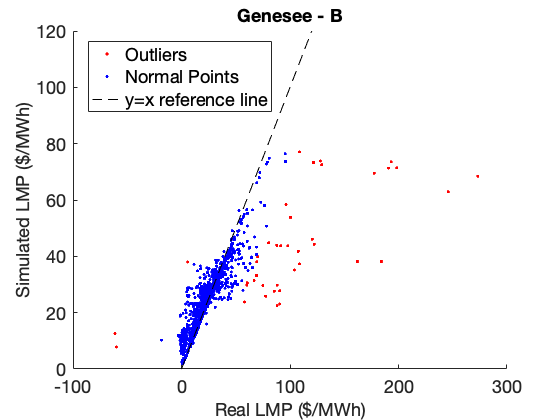}
         \caption{Winter LMP comparison for zone B}
         \label{fig:winter}
     \end{subfigure}
     \hfill
     \begin{subfigure}[b]{0.49\columnwidth}
         \centering
         \includegraphics[width=\textwidth, trim = {0.5cm 0 1cm 0},clip]{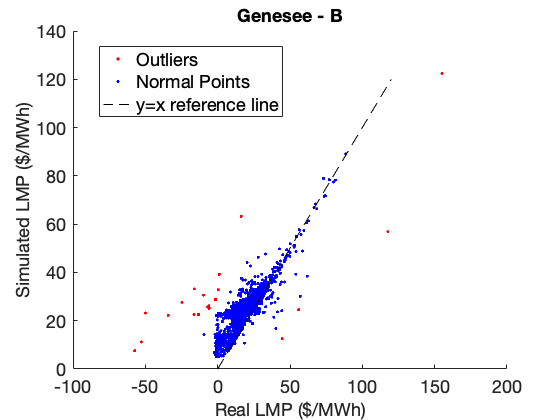}
         \caption{Summer LMP comparison for zone B}
         \label{fig:summer}
     \end{subfigure}
     
        \caption{LMP comparison for 5pm on Dec 19, 2019}
        \label{fig:lmpcompB}
\end{figure}

\begin{figure}
     \centering
     \begin{subfigure}[b]{0.49\columnwidth}
         \centering
         \includegraphics[width=\textwidth,trim = {0.5cm 0 1cm 0},clip]{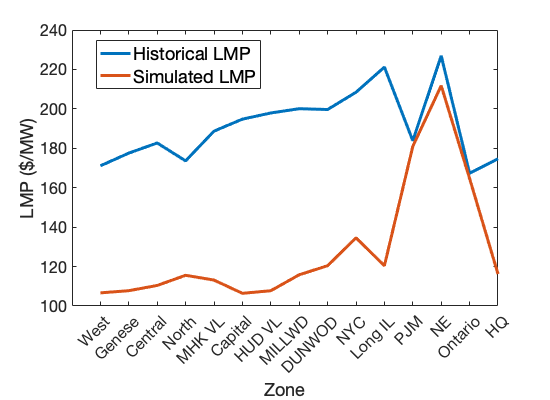}
         \caption{LMP comparison without modification}
         \label{fig:wintercase_a}
     \end{subfigure}
     \hfill
     \begin{subfigure}[b]{0.49\columnwidth}
         \centering
         \includegraphics[width=\textwidth, trim = {0.5cm 0 1cm 0},clip]{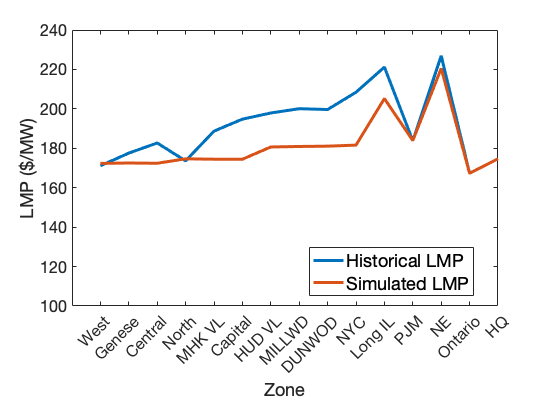}
         \caption{LMP comparison with scaled up fuel price}
         \label{fig:wintercase_b}
     \end{subfigure}
     
        \caption{LMP comparison for 5pm on Dec 19, 2019}
        \label{fig:wintercase}
\end{figure}

Figure \ref{fig:lmpcompB} shows the real and simulated LMPs comparison for zone B, where the x-axis indicates the real LMPs and the y-axis indicates the simulated LMPs. We draw the $y=x$ line as a reference for the perfect match of the prices. Zone B is chosen because it has relatively simple thermal generator compositions as shown in Figure \ref{fig: overview}, meaning the dispatch strategies are more likely to be straightforward. We define the outliers as points that are $\pm30\$\/MWh$ off the reference price. We choose this threshold to limit the number of outliers to an analyzable size and focus on the extreme cases. 43 outliers are identified for winter and 27 are identified for summer, indicated by the red dots. The outliers in winter mostly happened at peak (or near peak) hours of extremely cold days (Jan 21-22 and Dec 18-22), aligning with the fuel price peaks highlighted in the NYISO winter 2019-2020 Cold Weather Operations report \cite{Yeomans2020}, which were approximately five times higher than normal.\\
Figure~\ref{fig:wintercase_a} compares the simulated and real prices across the system at 5pm on December 19 as an example. Without modifications, the simulated LMPs (orange line) do not accurately represent the historical LMPs (blue line). However, when fuel price is scaled by a factor of five to more closely represent the real price peak, the comparison improves, as shown in in Figure~\ref{fig:wintercase_b}. By adjusting the fuel price, the fit of simulated LMPs improve significantly, suggesting that improvements in fuel price prediction, and generator bidding strategy in general, can improve the simulated LMP performance.

\begin{figure}
     \centering
     \begin{subfigure}[b]{0.49\columnwidth}
         \centering
         \includegraphics[width=\textwidth, trim = {0.5cm 0 1cm 0},clip]{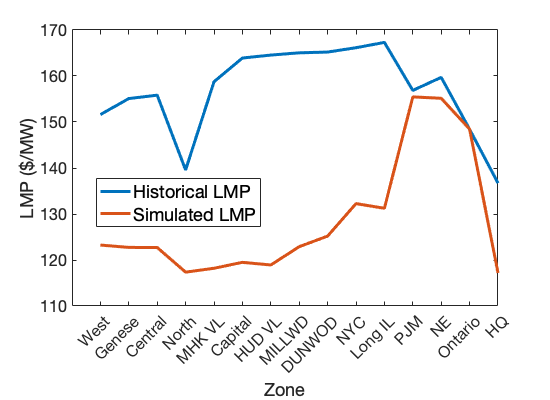}
         \caption{LMP comparison without modification}
         \label{fig:summercasea}
     \end{subfigure}
     \hfill
     \begin{subfigure}[b]{0.49\columnwidth}
         \centering
         \includegraphics[width=\textwidth, trim = {0.5cm 0 1cm 0},clip]{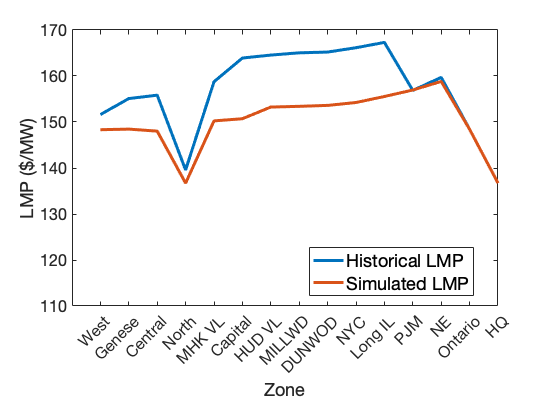}
         \caption{LMP comparison with scaled up fuel price}
         \label{fig:summercaseb}
     \end{subfigure}
     
        \caption{LMP comparison for 6pm on Aug 3, 2019}
        \label{fig:summercase}
\end{figure}

\vspace{-2pt}
A similar analysis for summer season shows two types of outliers. One type is similar to the winter case, which is related to hot weather and peak loads (shown in red on the right side of the reference line in Figure~\ref{fig:summer}). The other type is attached to very low or even negative real LMPs. The negative real LMPs are caused by the negative bidding price \cite{nyisodata} which is not modeled by the cost curve of generators. According to the NYISO's 2019 market report \cite{Monitor2019}, the weather in summer 2019 was mild and thus does not exhibit many outliers caused by high load. The simulation result of one such data point 6 pm on Aug 3 is shown in Figure \ref{fig:summercasea}. The simulated LMPs for all load zones are significantly lower than the real LMPs. From \cite{nyisodata}, the ancillary services price was very high for the examined hour. With the fuel price scaled up by 50\% to approximate the potentially high price for reserves, the result shown in Figure \ref{fig:summercaseb} illustrates improvement in the ability of the simulated LMPs to capture overall trend and magnitude of the real LMPs.

\section{Conclusion and Discussion}\label{sec:conclusion}
In this paper, we proposed a baseline model for NYS using only publicly available data sources. A modified WARD network equivalent reduction method is used to simplify the NPCC 140-bus model, which represents the backbone of the northeast region of Eastern Interconnection. The reduced network is then further modified to more closely reflect the current NYS transmission system. Load and generation profiles are updated for the NYS area with data from 2019 to provide model validation and to support future research. The PF validation indicates that with precise records of generation profiles, the baseline model can approximate the major interface flows with 90\% accuracy for most hours in 2019. A comparison of LMPs further demonstrates that the baseline model is capable of representing the major characteristics of the real grid, with high correlation under common situations. With modification of cost curves to represent peak conditions, the simulation results are significantly improved. The baseline model is coded in MATPOWER with interfaces to access and compile raw data from online resources, and is publicly available on GitHub. This open-source access provides other users the ability to  adjust the generator/load profiles or make transmission network updates in the future for different purposes.
 
As previously described, the bidding strategy of generators, hydro power operations, long-term contracts with external areas and non-flexibility of reserve resources are not considered in the baseline model. However, the importance of precise hydro generation profile and its impact on PF performances is discussed. Specifically, the accuracy of simulated LMPs suffers in zone A as a function of Robert Moses Niagara hydro bidding strategies that depend on multiple factors that are not included in this model (such as the availability of the water, environmental regulations etc). Exploring the bidding strategies of generators and incorporating a reserve market would be a valuable research direction to further improve this baseline. It is also worth noting that correlation coefficients between the simulated and real LMPs are lower when the composition of thermal generators becomes increasingly complex. The non-flexible reserve resources in J and K require additional attention as more intermittent renewable energy is integrated in the near future.
 
In summary, the baseline model proposed and validated in this paper is intended to be a tool for researchers, stakeholders, and policymakers to support analysis and understanding of the current NYS network. We hope that this model will be leveraged to answer a variety of research questions on this open-source, customizable model and dataset. In particular, with the geospatial information encoded in each bus, the model could serve as a powerful underlying foundation for distributed and intermittent renewable integration studies that are sensitive to spatial and temporal correlations.

\section*{Acknowledgment}

BY, JAS and KMZ would like to acknowledge the funding support from the New York State Energy Research and Development Authority (NYSERDA). JAS would also like to acknowledge his support by the National Science Foundation Graduate Research Fellowship Program under grant number DGE-1650441. CLA and MVL would like to acknowledge the support of USDA National Institute of Food and Agriculture (NIFA) under Grant No. 2019-67019-30122.

\ifCLASSOPTIONcaptionsoff
  \newpage
\fi



%
\bibliographystyle{ieeetr}
\bibliography{reference}

\end{document}